\documentclass[12pt,preprint]{aastex}

\newcommand\HI{\hbox{H\footnotesize I}}
\newcommand\HII{\hbox{H\footnotesize II}}
\def\simlt{\lower.5ex\hbox{$\; \buildrel < \over \sim \;$}}
\def\simgt{\lower.5ex\hbox{$\; \buildrel > \over \sim \;$}}

\slugcomment{ApJ,in press}

\shorttitle{Ionizing background and the $N_{HI}$ distribution}
\shortauthors{Corbelli and Bandiera}

\begin{document}


\title{The effects of an ionizing background on the \HI \\ 
       column density distribution in the local Universe}

\author{Edvige Corbelli and Rino Bandiera}
\affil{Osservatorio Astrofisico di Arcetri, L. E. Fermi, 5, 50125
    Firenze, Italy}
\email{edvige@arcetri.astri.it, bandiera@arcetri.astro.it}

\begin{abstract}
Using data on the \HI\ column density distribution in the local Universe, 
$f(N_{HI})$, in this paper we show how to determine $g(N_{H})$, 
the distribution of the total gas (\HI+\HII) column density. A simple 
power law fit to $f(N_{HI})$ fails due to bendings in the distributions  
when $N_{HI}<10^{20}$~cm$^{-2}$ and H is no longer fully neutral. 
If an ultraviolet background  
is responsible for the gas ionization, and $g(N_{H})\propto N_{H}^{-\alpha}$, 
we find the values of $\alpha$ and of the intensity of the background 
radiation which are compatible with the present data. The best fitting values of 
$\alpha$, however, depend upon the scaling law of the the gas volume densities 
with $N_{H}$ and cannot be determined unambiguously. We examine in detail 
two models: one in which the average gas volume density decreases  
steadily with $N_H$, while in the other it stays constant at low column 
densities.
The former model leads to a steep power law fit for $g(N_{H})$,
with $\alpha\simeq 3.3\pm 0.4$ and requires an ultraviolet flux larger 
than what the QSOs alone produce at $z=0$. For the latter 
$\alpha\simeq 1.5\pm 0.1$ and a lower ionizing flux is required. 
The ambiguities about the modelling and the resulting
steep or shallow $N_{H}$ distribution can be resolved only if
new 21-cm observations and QSOs Lyman limit absorbers searches 
will provide more
data in the \HI-\HII\ transition region at low redshifts. 
Using the best fit obtained for higher redshift data we outline  
two possible scenarios for the evolution of gaseous structures, 
compatible with the available data at $z\sim 0$. 

\end{abstract}


\section{Introduction}

The \HI\ column density distribution of intergalactic gas clouds, $f(N_{HI})$,
is determined at high redshifts through the analysis of QSO absorption spectra.
The detection of Lyman continuum absorption and of Ly$\alpha$ absorption lines
between us and the QSOs has been used to derive $f(N_{HI})$ from 
$N_{HI}\simeq 10^{12.5}$~cm$^{-2}$ to
$N_{HI}\simeq 10^{21.5}$~cm$^{-2}$ \citep[e.g.][]{pet93,rau98,sto00}.
Attempts have been made to fit $f(N_{HI})$ over many decades of $N_{HI}$
with a simple power law.
However, it is more physically meaningful to expect a power law behavior for
$g(N_{H})$, the total (neutral+ionized) column density distribution. In fact,
due to drastic changes in the hydrogen ionization fraction when the gas becomes
optically thin to photons above 13.6~eV, breaks in $f(N_{HI})$ appear for
$N_{HI}\sim 10^{18}$--$10^{20}$~cm$^{-2}$, where a small decrease of $N_{H}$
corresponds to a rapid change in $N_{HI}$.
A more appropriate approach is to compare the observed $f(N_{HI})$ to the $N_{HI}$
distribution function derived from an assumed $g(N_{H})$, after ionization
changes corrections have been applied. This method has been used by
\citet{cor01} to derive the slope of $g(N_{H})$,
assumed to be $\propto N_H^{-\alpha}$, and the ionization conditions of the
gas at redshifts $z\sim 2$--$3$.
Since the frequency of intervening absorption systems is high at these
redshifts, it has been possible to limit the analysis to high column density
absorbers, using data relative to (mostly neutral) Damped Ly$\alpha$ absorption
systems, and to (mostly ionized) Lyman limit systems.
A comparison of models for dark matter confined systems with the observed $f(N_{HI})$ 
led to values of $\alpha > 2$ for $z\sim 2$--$3$; an extrapolation of this power law
distribution towards systems of lower column density, the Ly$\alpha$ forest,
gives reasonable fits for $N_{HI}>10^{15}$~cm$^{-2}$.

At lower redshifts the number density of absorption systems is much lower and
the poor statistics based on Lyman limit and Damped Ly$\alpha$ alone does not
allow a determination of the distribution function $g(N_{H})$.
At zero redshift however these studies can be complemented with detailed
observations of 21-cm emission line radiation from nearby galaxies.
21-cm emission data have the advantage of leaving small uncertainties on
$N_{HI}$, while saturated Lyman limit breaks in QSOs spectra give only lower
limits on the intervening $N_{HI}$ \citep[e.g.][for a correct statistical
treatment of such measurements]{ban01}.
If 21-cm observations are sensitive enough to \HI\  column densities where the
hydrogen gas is mostly ionized in the presence of a UV radiation field,
then we could  use these data to infer $g(N_{H})$ and the intensity of the
ionizing background radiation in the local Universe, which is not directly
observable. Similarly, the observed sharp drop of the \HI\ column
density in the outer disk of two nearby spiral galaxies, M33 and NGC3198, 
interpreted as an \HI-\HII\ transition zone, made possible an estimate 
of the UV ionizing flux at $z=0$ \citep{cor93,mal93}. The known 
dark matter content of these galaxies, inferred from the gas kinematics, 
reduced the associated uncertainties. 
In this paper we shall give additional evidence for a UV ionizing background
in the local Universe using the shape of the observed $f(N_{HI})$. 
We derive power law fits to $g(N_{H})$ which reproduce $f(N_{HI})$
quite well over many orders of magnitude, after ionization corrections 
have been applied.
We underline the observations needed in the future to reduce the
ambiguities in modeling the ionization corrections and the
resulting best fitting values of $g(N_H)$. We briefly discuss two different 
evolutionary scenarios for gaseous structures in the Universe, both compatible
with the present data.

\section{The $N_{HI}$ distribution function at $z=0$}

At low redshifts there are two different sets of data which can be used
to derive $f(N_{HI})$:

{\it - Ly$\alpha$ absorption lines and Lyman breaks observed in QSOs spectra}.
The Hubble Space Telescope (HST) has provided good quality data for low
redshift absorption systems.
\citet{pen00} and \citet{shu01} have analyzed HST/FUSE data for redshifts
$z<0.1$ and $N_{HI}< 10^{17}$~cm$^{-2}$.
\citet{wey98} have analyzed data for $N_{HI}< 10^{16.5}$~cm$^{-2}$ in a
broader redshift range ($z < 1.3$) and found a weak evolution in the number of
absorbers per unit redshift.
As \citet{pen00} have already pointed out there is good agreement between the
$N_{HI}$ distributions derived from the two sets of data and therefore we limit
ourselves to the \citet{wey98} data, shown in Figure~1$(a)$.
For higher column densities $f(N_{HI})$ at $z\approx 0$ is
evaluated from the Lyman limit and Damped Ly$\alpha$  data collection of
\citet{ban01} over the redshift interval $0\le z < 1.3$, taking into account 
the mild evolution of Lyman limit systems \citep{cor01}.

{\it - 21-cm maps of nearby galaxies and the \HI\  mass function}. 
21-cm observation of the \HI\  gas in nearby galaxies with high resolution and
sensitivity provide data on $d\Sigma(N_{HI})$, the galaxy differential cross 
section for a given range of $N_{HI}$, averaged over all inclinations.
This information, acquired for galaxies of different \HI\  masses and coupled
with the \HI\  mass function, $\Phi(M_{HI})$, gives $f(N_{HI})$:

\begin{equation}
f(N_{HI})= {\int_{M_{HI}^{min}}^{M_{HI}^{max}}
\Phi(M_{HI}) d\Sigma(N_{HI},M_{HI}) dM_{HI}\over 
(H_0/c)\ dN_{HI}} 
\end{equation}

\noindent
In order to determine $f(N_{HI})$ across the region where the hydrogen becomes
mostly ionized, observations should be sensitive to column densities down
to $\sim 3\times 10^{18}$~cm$^{-2}$.
Recently the \HI\  Parkes All Sky Survey (HIPASS) together with the Australia
Telescope Compact Array (ATCA) has provided good quality data for
$\Phi(M_{HI})$ and for $d\Sigma(N_{HI},M_{HI})$ \citep{rya01}. Fitting the data
with a Schechter function \citep{sch76},
the resulting $\Phi(M_{HI})$ has a characteristic \HI\ mass value  
$M^*=1.14\times 10^{10}$ M$_\odot$, a faint end slope
$\alpha=1.52$, and a normalization $\theta=0.0032$ \citep{kil99,kil00}.
Similar values have been obtained recently also by \citet{ros01}.
We use this mass function in the following analysis and consider 6 mass bins of
equal width across the range $7.5<\log(M_{HI}/{\hbox{M}}_\odot$)$<10.5$.
\citet{rya01} have evaluated $d\Sigma(N_{HI},M_{HI})$ for two $M_{HI}$ ranges: for
$10^{7.5}<M_{HI}<10^8$~M$_\odot$ and for $10^{10}<M_{HI}<10^{10.5}$~M$_\odot$.
In the literature there are furthermore two galaxies which have been mapped
in great detail down to a low surface density: M33 and NGC3198. We compute 
$d\Sigma$ for these galaxies as orientation averages of the face on cross
sections,  obtained from the deconvolution of the observed $N_{HI}$ radial
profiles with the best fitted tilted ring model \citep{beg89,mal93,cor93,cor00}.
The \HI\ mass of M33 is $\sim 2.7\times 10^{9}$~M$_{\odot}$ for a distance of
0.84~Mpc and that of NGC3198 is $\sim 5\times 10^9$~M$_{\odot}$ for a distance
of 9.4~Mpc.
Data on these two galaxies can therefore be used for $d\Sigma(N_{HI},M_{HI})$
in the two mass bins spanning $10^{9}<M_{HI}<10^{10}$~M$_{\odot}$.
In order to estimate $d\Sigma$ for $10^{8}\le M_{HI}\le
10^9$~M$_{\odot}$, where we have no data, we have averaged the values 
on the adjacent $M_{HI}$ bins.
$f(N_{HI})$, as derived from equation~(1), is plotted in Figure~1$(b)$.
The ATCA data \citep{rya01}, as well as the M33 and NGC3198
data, show the ``footprint'' of ionization.
While in 21-cm emission maps the ionization has the effect of giving a sharp
drop to the radial decline of $N_{HI}$, in the distribution function this
translates into a plateau where the \HI-\HII\ transition occurs.
When the gas becomes mostly ionized, at lower $N_{HI}$ values, $f(N_{HI})$
rises again but it is offset with respect to the neutral side and has a slope 
which depends on how the \HI/\HII\ ratio varies with $N_{H}$ \citep{cor01}.
Figure~1$(b)$ shows that this is indeed the behavior of $f(N_{HI})$, even
though the sum over different mass bins may have smoothed somewhat the
\HI-\HII\ transition (if the gas vertical stratification depends on
the galaxy mass).

\section{From the shape of $f(N_{HI})$ to the total gas distribution}

In order to evaluate $f(N_{HI})$ over the range
$10^{14}\le N_{HI}< 6\times 10^{20}$~cm$^{-2}$ in the local Universe
we select QSOs absorption data for $N_{HI}< 5\times 10^{17}$~cm$^{-2}$ and
21-cm emission data at higher $N_{HI}$.
We exclude both the very low column density Ly$\alpha$ forest, since it may
not be associated with denser structures, and the very high column density data,
because star and molecular gas formation might have reduced the original
\HI\ column density.
While simple power law fits to $f(N_{HI})$ fail we will show in what follows
that a power law for the total hydrogen distribution can instead well fit all
the data displayed in Figure~2.
As in \citet{cor01} we derive a theoretical $f(N_{HI})$ after applying the
$N_H-N_{HI}$ conversion relation to an assumed power law distribution for the
total gas column density, $g(N_{H})=K N_{H}^{-\alpha}$.
The use of 21-cm data eliminates the problem of having large errors in $N_{HI}$
in the region where Lyman limit systems become saturated and therefore we can
safely use $f(N_{HI})$ binned over small $N_{HI}$ intervals.
We consider two possible $N_{H}-N_{HI}$ conversion models for
$10^{14}\le N_{HI}<6\times 10^{20}$~cm$^{-2}$:

-$(a)$ the model used by \citet{cor01} to fit the data on Lyman limit and 
Damped Ly$\alpha$ systems, as described in their Section 3, extended to lower
column density systems.
This model is valid for a dark matter confined gas with ionization fractions
which increase steadily as the total gas column density decreases.

-$(b)$ the model used by \citet{cor01} only for $N_{H}\ge N_*$; for $N_{H} < N_*$
we keep constant the H ionization fraction, equal to that reached for
$N_{H}=N_*$.
A constant ionization fraction for $N_{H}< N_*$ is achieved if the average gas
volume density stays constant as $N_{H}$ decreases below $N_*$.
A constant gas volume density might imply either that a constant external
pressure is confining the gas or that dark matter and turbulence are
distributed in such a way as to give a gas vertical dispersion which does not
depend on the gas column density.

For both $(a)$ and $(b)$ we find the best fitting value of $\alpha$ and of
$J_L/\eta_0$. This is the ratio between the intensity of the background flux at 912
\AA\ normalized to $10^{-22}$~ergs~cm$^{-2}$~s$^{-1}$ Hz$^{-1}$~sr$^{-1}$, and
the gas compression factor $\eta_0$ due to gravity from matter other than
the gas (e.g.dark matter). 
The constant $\eta_0$ adds to the gas self gravity in vertical scale height 
expression \citep{cor01}:

\begin{equation}
{1\over h^2}={16 G^2 M_g^2\over c_s^4} \left(1 + \eta_0^2 {c_s^2 \over 
\tilde c_s^2 }\right) 
\end{equation}

where $M_g$ is the total mass surface density per cm$^{-2}$ of gas,
$c_s$ is the thermal sound speed and $\tilde c_s$ is the value
of $c_s$ for $T=10^4$ K. 
For model $(a)$ the minimum value of $\chi^2$ (1.3) corresponds to
$\alpha=3.32$ and to $\log J_L/\eta_0=-0.25$.
The 3-$\sigma$ uncertainties on $\alpha$ are $\pm 0.4$, on $\log J_L/\eta_0$
are $\pm0.6$.
An $\alpha=3$ power law is therefore compatible with the data. This is
important since there have been often claims that $f(N_{HI})$ for disk like
structures should follow an $N_{HI}^{-3}$ power law \citep{mil88}, but
$f(N_{HI})$ observed in galaxies strongly deviates from this law
\citep{zwa99}. This is not surprising since it is $g(N_{H})$ which  
should eventually scale as $N_{H}^{-3}$ for disk like systems, and 
we have shown that for model $(a)$ the present data are compatible with 
this law. For model $(b)$ the best fit gives $\alpha=1.47\pm0.1$, $\log
J_L/\eta_0=-1.85\pm 1 $, and $\log N_*/{\hbox{cm}}^{-2}= {19.4\pm 0.4}$.
This model fits the data better than model $(a)$ with 0.5 being the minimum
$\chi^2$ value.

The resulting $f(N_{HI})$ are shown in Figure~2 together with the data used 
for the $\chi^2$ minimization.
Both best fitting models reproduce rather well also the high and low column
density QSO's absorption data, not used for the fit, but shown in the 
Figure as well. A pure power law with no 
ionization corrections fails since it gives a minimum $\chi^2\sim10$.
Unfortunately the present data are not sufficient to discriminate between 
$(a)$ and $(b)$ type of model.
This is mostly due to the lack of data around $N_{HI}\simeq
10^{18}$~cm$^{-2}$ and to the uncertainties on the conversion of atomic 
hydrogen gas into other baryonic forms at very high column densities.
In these two regions $(a)$ and $(b)$ model predictions for $f(N_{HI})$ 
differ substantially.

The two best fitting power laws for $g(N_{H})$ imply a quite different degree of
ionization and therefore different total gas densities in the local
Universe.
For model $(a)$ most of the gas resides in low column density absorbers; 
H neutral fractions are of order 0.002 for $N_{HI}\simeq 10^{17}$~cm$^{-2}$ and
are as low as $10^{-5}$ for $N_{HI}\simeq 10^{14}$~cm$^{-2}$.
The total gas density(H+He) predicted by this distribution 
for $10^{14}\le N_{HI}\le 10^{22}$ gives $\Omega_g h_{50} (z=0) \simeq 0.01$ 
with a 3$\sigma$ range of $0.005-0.025$.
Model $(b)$ instead predicts a lower total density of gas in the same range
of $N_{HI}$: $\Omega_g h_{50} (z=0) \simeq 10^{-3}$ with a 3$\sigma$ range
of $0.001-0.005$.

\section{Discussion}

We now discuss the best fitting power laws for $g(N_{H})$ derived in the
previous Sections, taking into account the known limits on the UV ionizing
background flux at $z=0$, the limits on the density of baryonic matter,
$\Omega_b$, and the total gas column density distribution obtained for QSOs
absorbers at $z\sim 2 - 3$ by \citet{cor01}.
Limits on the intensity of the UV ionizing background at $z=0$ rely on several
indirect techniques such as those based on the  detection of a sharp cutoff in
the \HI\ surface density of outer disks of nearby galaxies
\citep{cor93,mal93,dov94}, on searches for H$\alpha$ emission from outer disks
or from intergalactic \HI\  clouds \citep{don95,mad01,wey01}, on measurements of
column densities of metal species in intergalactic \HI\ clouds \citep{tum99}.
The range $J_L\simeq 0.1-0.4$ is consistent with the above limits as well as
with the determination of the UV ionizing background from the QSOs luminosity
function, even considering possible additional contributions of photons escaping  
from star forming galaxies \citep{har96,shu99}.
Measurements of the proximity effect for $z<1$ have been reported
\citep{kul93,pas01,sco01} but the resulting ionizing background intensity,
$J_L\simeq 0.01-4$, is highly uncertain due to possible clustering of the
Ly$\alpha$ forest around QSOs and to a possible evolution over the used
$z$ range.
The total density of baryonic matter can be inferred from deuterium abundances,
$\Omega_b h^2_{50}\simeq 0.02$ \citep{bur01,ome01}.
For a Universe with $\Omega_\Lambda=0.7$ and $\Omega_M=0.3$ this $\Omega_b$
value limits the power law index of the total gas column density distribution
derived by \citet{cor01} at higher redshifts to values $\alpha < 3$. The
extrapolation of the same power law towards the
Ly$\alpha$ forest requires however that ionization conditions of the gas stay 
constant for $N_{HI}< 10^{15}$~cm$^{-2}$ at these redshifts 
i.e.\ for $N_{H}\simlt 2.5\times 10^{19}$~cm$^{-2}$.
In the following we shall assume that the gas distribution at $z\sim 2 - 3$ is
well described by the Best-Fit 1 model of \citet{cor01} ($\alpha=2.7$,
$\eta_0=12.5$, $\log J_L/\eta_0=-0.35$, see their Table 1).

If today the gas is distributed according to the best fitting model $(a)$, with
$\alpha \simeq 3.3 $, the limits on $J_L$ and on $\Omega_b$ at $z=0$ force the
gas compression factor to be $\eta_0<3$.
According to this model the following evolutionary scenario might hold: because
changes in time of the ionization conditions of the gas are mild, 
the average gas compression factor should decrease with time in order to balance 
the decrease of $J_L$. If the dark matter is confining and compressing the gas
then today's systems have on average a lower dark matter density than systems at
earlier times.
The alternative would be to invoke a steadily increasing turbulence which 
lowers the effective compression $\eta_0$, or additional ionization sources in
the surroundings of gaseous clouds.
The power law index for the total gas column density 
distribution from $z\sim 2.5$ to $z\sim 0$ becomes slightly steeper and now 
it is compatible with a disk-like geometry. For model $(a)$ the gas associated 
with the Ly$\alpha$ forest clouds is, even
today, the dominant reservoir of baryonic matter in the Universe.

The best fitting model $(b)$ at $z\sim 0$ implies a constant hydrogen neutral
fraction, of order $0.05$, for $N_{HI}\simlt 10^{18}$~cm$^{-2}$.
The power law index for $g(N_{H})$ in this case is $\alpha\simeq
1.5$ and $J_L/\eta_0\simeq 0.01$.
In order to explain the flattening of the power law index for $g(N_{H})$ with
time, we invoke a scenario in which low column density
structures collapse towards the center or merge to form higher column density
structures.
If $\eta_0$ is constant through the evolution of the Universe, the required
ionizing background radiation at $z=0$ would equal that estimated by
\citet{har96} for QSOs as primary sources of ionizing photons.
If the external pressure is responsible for confining the gas and keeping
constant the ionizing fraction at low column densities, the inferred value of
$P_{ext}$ is also constant through redshifts and is
$P_{ext}$/(cm$^{-3}$~K)~$\simeq \eta_0\simeq 10$~\citep{cha94,cor95}.
This excludes that $P_{ext}$ is dominated by a diffuse and pervasive intergalactic 
medium since this is expected to decrease with time.
However, the merging or the collapse of low column density structures is likely
to trigger star formation and outflows of hot gas into the intergalactic
medium; this hot gas at temperatures T$\sim 10^5$--$10^7$~K would be the likely
confining medium of low column density Ly$\alpha$ clouds and it would
contribute mostly to the density of baryonic matter in the Local Universe
\citep{fuk98,cen99,dav01,tri01}.

In this paper we have shown that ionizing radiation affects 
the neutral gas distribution in the local Universe
and that any attempt to fit $f(N_{HI})$ with a simple power law across the
\HI-\HII\ transition region ($10^{17}<N_{HI}<10^{20}$~cm$^{-2}$) can lead to
misleading results. A power law distribution for the total gas column
density instead fits the data reasonably well, from the Ly$\alpha$
forest to the high gas column density in galaxies.  
Additional 21-cm data, as well as new Lyman limit searches at low
redshifts, are needed to implement the knowledge of $f(N_{HI})$ in this delicate 
region and draw a more definite conclusion on the slope of the
total gas column density distribution.
Furthermore, if 21-cm data are analyzed in detail to constrain the gas and
dark matter distribution in outer disks, one can assign a more definite value to
the gas compression and picture a unique evolutionary scenario for the
gas distribution and the UV ionizing background in the Universe.

\acknowledgments

We are grateful to the anonymous referee for a careful reading of the original 
manuscript.

\clearpage


\begin{figure}
\plotone{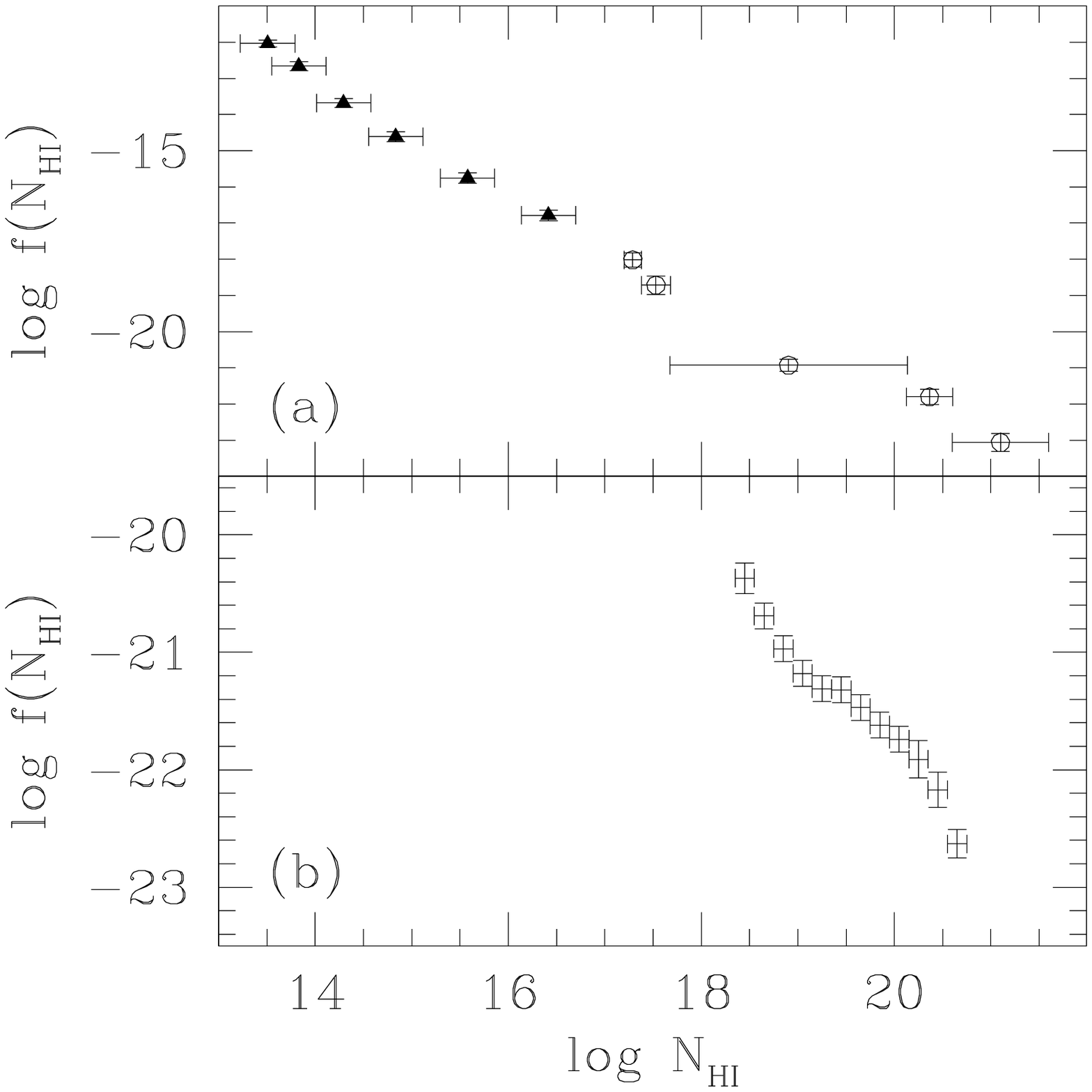}
\caption{
$(a)$ $f(N_{HI})$ from QSO Ly$\alpha$ and Lyman continuum absorption data at
low redshifts.
Ly$\alpha$ forest data from \citet{wey98} are shown as filled triangles while
data for Lyman limit and Damped Ly$\alpha$ systems from our collection are
shown as open dots.
$(b)$ $f(N_{HI})$ from 21-cm data at $z=0$ (see text for details).
\label{fig1}}
\end{figure}

\begin{figure}
\plotone{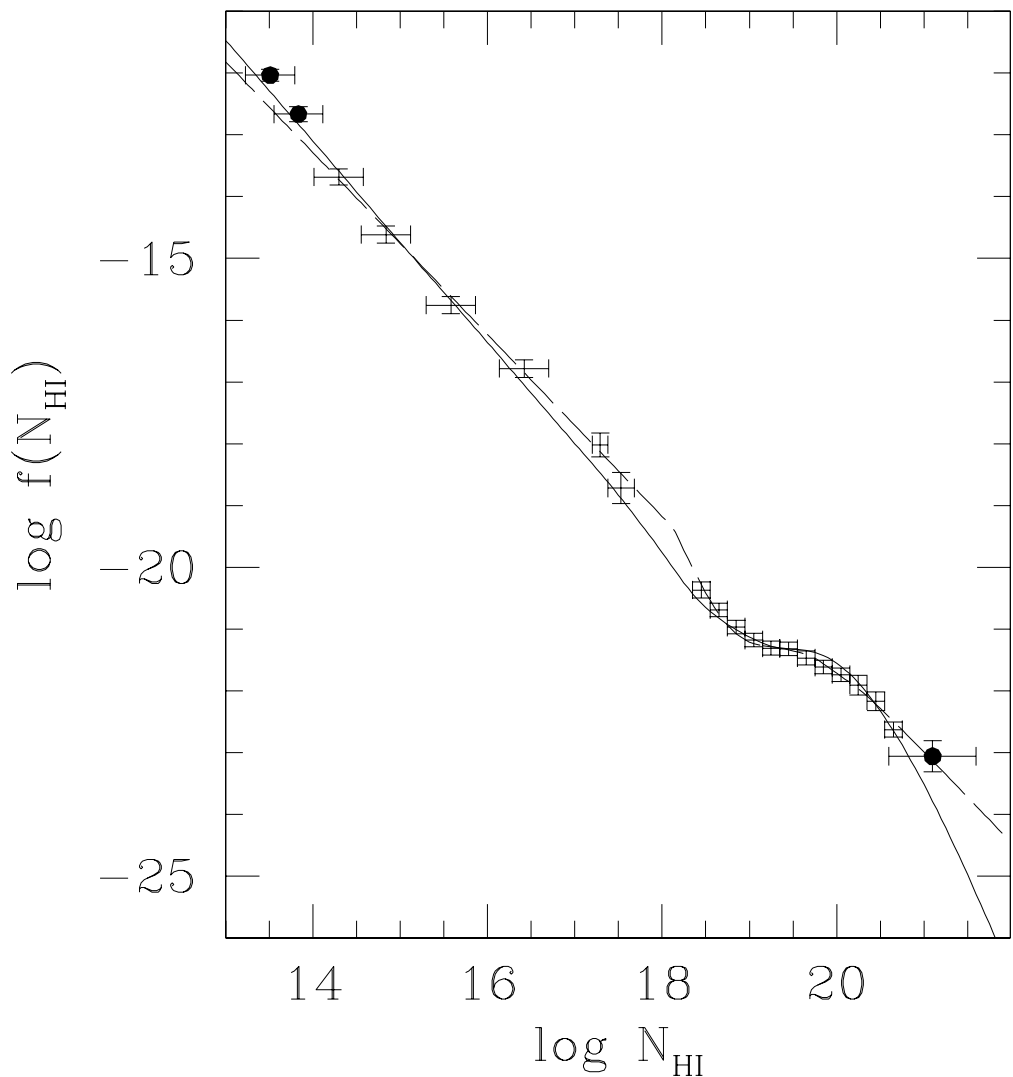}
\caption{
The two theoretical $f(N_{HI})$ which best fit the data for
$10^{14}<N_{HI}<6\times 10^{20}$~cm$^{-2}$, shown in the Figure with simple crosses.
The continuous line is derived from a total gas distribution function with
$\alpha=3.32$ using the ionization model $(a)$ with $\log J/\eta_0=-0.25$.
The dashed line is derived from a total gas distribution function with
$\alpha=1.47$ using the ionization model $(b)$ with $\log J/\eta_0=-1.9$ and
$N_*= 2.5\times 10^{19}$~cm$^{-2}$.
The high and low column density data shown with filled dots have not
been used for the fit.
\label{fig2}}
\end{figure}

\end{document}